\documentclass[preprint2]{aastex}

\begin{document}

\shortauthors{Struck and Smith}
\shorttitle{Models of Arp 284}


\title{Models of the Morphology, Kinematics, and Star Formation
History of the Prototypical Collisional Starburst System: NGC 7714/7715
= Arp 284}

\author{Curtis Struck}
\affil{Department of Physics and Astronomy, Iowa State University,
    Ames, IA 50011}

\author{Beverly J. Smith}
\affil{Department of Physics and Astronomy, East Tennessee State
Univ., Box 70652, Johnson City, TN 37614}


\begin{abstract}

We present new N-body, hydrodynamical simulations of the interaction
between the starburst galaxy NGC 7714 and its post-starburst companion
NGC 7715, focusing on the formation of the collisional features,
including: 1) the gas-rich star forming bridge, 2) the large gaseous
loop (and stellar tails) to the west of the system, 3) the very
extended HI tail to the west and north of NGC 7714, and 4) the partial
stellar ring in NGC 7714.  Our simulations confirm the results of
earlier work that an off-center inclined collision between two disk
galaxies is almost certainly responsible for the peculiar morphologies
of this system.  However, we have explored a wider set of initial
galaxy and collisional encounter parameters than previously, and have
found a relatively narrow range of parameters that reproduce all the
major morphologies of this system.

The simulations suggest specific mechanisms for the development of
several unusual structures.  We find that the complex gas bridge has
up to four distinct components, with gas contributed from two sides of
NGC 7715, as well as from NGC 7714.  The observed gas-star offset in
this bridge is accounted for in the simulations by the dissipative
evolution of the gas. The models suggest that the most recently formed
gas bridge component from NGC 7715 is interacting with gas from an
older component.  This interaction may have stimulated the band of
star formation on the north side of the bridge.  The models also
indicate that the low surface brightness HI tail to the far west of
NGC 7714 is the end of the NGC 7715 countertail, curved behind the two
galaxies. The sensitivities of the tidal structures to collision
parameters is demonstrated by comparisons between models with slightly
different parameter values.

Comparison of model and observational (HI) kinematics provides an
important check that the morphological matches are not merely
fortuitous. Line of sight velocity and dispersion fields from the
model are found to match those of the observations reasonably well at
current resolutions.

Spectral evolutionary models of the NGC 7714 core by Lan\c{c}on et
al. suggest the possibility of multiple starbursts in the last 300
Myr. Our hydrodynamic models suggest that bursts could be triggered by
induced ring-like waves, and a post-collision buildup of gas in the
core of the galaxy.

\end{abstract}

%

\section{Introduction}

Detailed investigations of individual collisional galaxies can provide
important information about the physics of the encounter, including
enhancement and suppression of star formation in interacting galaxies
and hydrodynamical processes such as shock wave production, heating
and cooling, and re-accretion.  In practice, however, it has proven
very difficult to reconstruct the details of a specific encounter
between galaxies, and to match precisely the observed properties of an
interacting system, particularly when one includes not just the
stellar morphology but also the gas distribution and kinematics and
the star formation morphology.

There are, however, some special circumstances in which such a
detailed modeling project is tractable.  The first of these is when
the galaxies are observed at a time shortly after closest approach,
and well before a merger.  In such cases, there has not been time for
phase mixing, and because most encounters are quite impulsive in the
early stages, the immediate response is primarily kinematic (e.g.,
\citet{ger93}, \citet{dub99}).  A second helpful factor is the
existence of a special symmetry in the collision. The symmetric
collisional ring galaxies provide a notable example \citep{app96}.
Another example is provided by the ocular galaxies, in which the
characteristic eye-like morphology results from a strong prograde
disturbance, implying that the companion orbital plane has a small
inclination relative to the disk plane of the ocular (\citet{elm91},
also see \citet{kau97}, \citet{kau99}, and \citet{elm00}).  Long tidal
tails also result from nearly planar prograde interactions
(\citet{too72}, and review of \citet{str99}), and can provide
important clues to the nature of an ongoing interaction.

The interacting system NGC 7714/15 (Arp 284), with a relatively nearby
distance of 37 Mpc (assuming H$_0$ = 75 km s$^{-1}$ Mpc$^{-1}$), has
many of these exceptional properties.  The most important features of
NGC 7714/15 are summarized in Tables l and 2, and are visible in the
optical \citet{arp66} Atlas photograph (Figure \ref{arp}, with a
Hubble Space Telescope archive image in Figure \ref{hst}), and in the
published 21 cm HI and H$\alpha$ maps (\citet{smi92}, hereafter Paper
I, and \citet{smi97}, hereafter Paper II).  NGC 7714 has a prominent
stellar ring \citep{arp66} which does not have on-going star
formation (\citet{bus90}; \citet{ber93}; \citet{gd95}; Paper II;
\citet{oha00}), or an HI counterpart (Paper II), unlike many
collisional rings \citep{app96}.  NGC 7714 has three apparent optical
tidal tails/arms.  The outer southwestern stellar arm may be
associated with a large HI loop (Feature 2 in Figure \ref{arp}, also
see Figure \ref{HImap}).  The inner southwestern arm (Feature 3 in
Figure \ref{arp}) is brighter, with a prominent HII region complex at
the base of this arm (Paper II).  In optical images, the northeastern
stellar arm (Feature 4 in Figure \ref{arp}) is clearly physically
separate from the bridge, however, in the HI maps, it is not
well-resolved (Paper II; see Figure \ref{HImap}).  In addition, low
resolution HI observations reveal a low surface brightness HI tail
extending 6$'$ (71 kpc) to the west of NGC 7714 with no observed
optical counterpart (Paper I).

The presence of the ring suggests a direct impact on the disk, while
the spirals suggest that the collision was somewhat off-center, and
generated significant tidal torques (Paper I).  The ring itself
constrains the collision parameters, but the spiral morphologies are
so distinctive that they should yield even stronger constraints. The
bridge consists of HI gas, old stars, and young stars, with a
significant offset between the old stars and the other components
(Paper II; see Figure \ref{HImap}).  This offset is clearly evident in
Figure \ref{hst}, where a string of luminous H~II regions lies to the
north of an optical continuum bridge.  Since models show that
connecting bridges are relatively short-lived features \citep{str97},
its presence provides strong evidence for a recent encounter. The
edge-on shape of NGC 7715 and the stellar countertail of NGC 7715 also
constrain the collision parameters.  Still more contraints are
provided by the map of line-of-sight HI velocities (Paper II), which
defines the sense of rotation of the two galaxies and kinematic line
of nodes.  NGC 7714 has a nuclear starburst (\citet{f80};
\citet{wee81}; \citet{k84}; \citet{gd95}; and \citet{kot01}), while
the center of NGC 7715 shows a post-starburst spectrum \citep{ber93}.
The stellar evolutionary timescales associated with these phenomena
give additional input on the time since the collision.

In Paper I, we presented a restricted 3-body model of the NGC 7714/5
encounter that matched the stellar morphology of the system and the
line of nodes and sense of rotation of NGC 7714.  This scenario
consisted of an off-center inclined collision between two unequal mass
galaxies, with the encounter being retrograde with respect to the main
galaxy NGC 7714 and prograde with respect to the companion.  This
simulation, however, did not include hydrodynamical effects.  The
later acquisition of the high resolution HI data (Paper II) allowed a
stronger test of the collision scenario.

In Paper II, we presented a preliminary hydrodynamical model of the
NGC 7714/5 encounter with two gas disks and rigid halos, using
parameters similar to those of the Paper I model.  This model was not
intended to reproduce all the features of the system, but rather to
demonstrate two points.  First, it showed that the gas loop could be
obtained in a collision like that in the model of Paper I with the
addition of a gas disk of larger radius than the stellar disk.
Second, it demonstrated that the observed offset in the old star
bridge versus the young star and gas bridge could be explained as a
result of the dissipative impact of the gas disks of the two galaxies.
This model did not, however, do a good job in matching the observed
location of the gaseous loop or the orientation of the gas/star offset
in the bridge.

In this paper we present the results of a more detailed modeling
program undertaken to better interpret the features listed in Tables 1
and 2.  The new modeling includes fully self-consistent simulations
made with the `Hydra' (version 3.0) N-body, adaptive-mesh SPH code of
\citet{cou95}, which includes radiative cooling, and also feedback
heating terms in some models (see the following section for details).
These new models are similar to the earlier ones in requiring a direct
impact between disks with a moderate impact parameter and a relatively
large inclination angle.  However, significant changes have been made
in the new models in order to better fit the detailed morphology of
the system (see section 3.1 and 3.2).  Comparisons to observed
kinematics (section 3.3) and star formation (SF) characteristics
(section 4) have also been made for the first time in this system.
The generally good agreement between models and observational
kinematics provides strong confirmation of the basic collisional
model.  The model results on SF are tentative, but they do provide
some useful suggestions to be checked in future work.  The results are
summarized in section 5.



\section{Numerical Models}

As noted above, the general nature of the collision can be immediately
deduced from the presence of a few distinctive morphologies.  The NGC
7714 ring and the relatively straight bridge connecting the two
galaxies provide prima facie evidence for a nearly head-on collision.
On the other hand, the offset of the NGC 7714 nucleus from the ring
center, and the presence of loops and tidal tails suggest some
asymmetry in the collision.  Tails and rings can be simultaneously
produced in collisions of intermediate inclination with closest
approach distances somewhat less than the radius of the gas disk of
the ring galaxy (e.g., \citet{app96}).  Tails are more easily produced
if the encounter is prograde relative to the tailed galaxy.

\subsection{Simulation Codes}

We began our modeling work with a large number of exploratory runs
with a restricted three-body code \citep{wal90}, in order to refine
our estimates of the collision parameters.  We will not describe that
work any further. We then used the hydrodynamic code of Paper II (see
\citet{str97} for details on this code) to study the gas dynamics of
the refined collision, and to further adjust it to reproduce the
observed HI structure. These runs were also used to study the thermal
and star-forming properties of the colliding galaxies in a preliminary
way.  However, that hydrodynamic code does not include fully
self-consistent calculations of the gravitational forces, and in
particular dynamical friction and related effects.  Thus, one of the
most prominent `errors' of typical models (e.g., models in which NGC
7715 begins nearly at rest relative to NGC 7714 at a distance of at
least several diameters away), is that NGC 7715 plunges through and
well away from NGC 7714, before the HI loop and other tidal structures
can develop to the observed degree. Given the limitations of these
models, we will not describe their results in any detail.

Fully self-gravitating simulations were then produced with the serial
code Hydra 3.0 (henceforth simply Hydra), which has been made publicly
available.  Hydra uses an SPH algorithm, and gravity is calculated
with an adaptive particle-particle (PP), particle-mesh ($AP^{3}M$)
algorithm (for details see \citet{cou95}, \citet{pea97}).  For a
typical timestep in our models, adaptive refinements were carried out
on about half the gas particles, with about 6-10 submeshes, and with
the most refinement around the primary center.

The simulations were all run using an adiabatic equation of state.
Optically thin radiative cooling was calculated via the tables of
\citet{sut93}, which were supplied with the Hydra code.  The
Sutherland and Dopita cooling curves include atomic and ionic line and
continuum processes for $T \ge 10^4 K$.  Cooling times were not used
to limit the size of the computational timestep, since the dynamical
time is usually longer than the cooling time.  Particles evolve
adiabatically and are cooled at the end of a given timestep, at
constant density. No feedback heating was included in most of the
Hydra simulations.

\subsection{Scalings and Boundary Conditions}

The codes used in this project run in dimensionless variables, and
many of the graphs below are plotted in those units. The Hydra runs
are made on a cubic volume, where x,y and z coordinates all run from
0.0 to 1.0 in code units.  Most of the computational cube is empty
most of the time (see Figures 1-3 below). Very few particles reach the
boundaries and when they do they are taken out of the calculation.
The adopted scalings for the Hydra model are: one computational length
unit equals $100 kpc$, one time unit equals $10^9 yr$, (which implies
that the velocity unit is 97.7 km/s), and the mass unit equals
$10^{10} M_\odot$.

With these scalings, the model results suggest that closest approach
occurred about $100-250 Myr$ ago; we will refine this estimate below.

\subsection{Initial Conditions and Model Differences}

We ran about a dozen simulations using the Hydra code; in this paper
we will focus on the results of the best Hydra simulation (henceforth
'the model' or 'best model'), with only brief mention of the results
of other model runs. In the best models the mass ratio of the galaxies
was about 1/3, in accord with the observations (e.g., the optical
luminosities). We note, however, that it is possible that the
companion has lost a good deal of mass, and thus, its precursor may
have been more massive than current observations suggest. A model with
equal mass initial galaxies showed that even with a massive companion
it is possible to produce a fairly good model, with only a few
disagreements with observation.

In the Hydra simulations, the two model galaxies each contain a
collisionless halo, a stellar disk, and a gas disk component, which
were added and relaxed sequentially.  The halo is approximately
isothermal, while the disk components have a nearly constant vertical
velocity dispersion with radius, and an approximately $1/r$ surface
density distribution.  The gas fraction of the disks is very high,
with equal masses of stars and gas.  This may be unrealistic, but was
done to provide adequate particle resolution of the gas dynamics.  The
gas disk cools somewhat in its relaxation, and so, is generally
thinner than the stellar disk.  Many details of the model galaxies are
given in Table 3.

In the models we adopt the x-y plane as the plane of the sky, and then
the model galaxy initial conditions and the orbital orbital
trajectories were optimized to match the observations at some later
time. Specifically, with the initial orientations given in Table 3 the
disk of the primary appears relatively face-on in the x-y plane, and
is edge-on in the x-z plane.  The companion disk is more nearly
face-on in the x-z plane.  Inner disk orientations are roughly
preserved through the collision.

Given the limited number of particles that can be used, we cannot
represent an extended halo with great accuracy with the Hydra code.
However, we find that the most successful models of this system
require relatively compact halos, which can be modeled quite well.

In the following sections, discussions of model results and
observational characteristics of the system will be highly interwoven.
To avoid confusion about which is being referred to we will refer to
the model galaxies as galaxies A and B, corresponding to NGC 7714 and
NGC 7715 respectively, and reserve the latter names for the real
galaxies.

\section{Model Results: Reconstructing the Collision}
\subsection {Overview of the Collision and Formation of the Large
Scale Morphology}

In this section, we will focus on the morphological and kinematic
results, and briefly discuss thermal effects and star formation in the
following section.  We begin with Figures \ref{gas1} and \ref{gas2},
which show the evolution of the gas disk of the model in three
orthogonal projections. The orbital trajectories of the galaxy centers
are also shown in the first row of Figure \ref{gas1}. The appearance
of the stellar disk during the last three time steps is shown in
Figure \ref{stars}.

In these figures, we see that galaxy B begins at some distance below A
in the z-direction. The galaxies swing around to almost return to
their starting positions. Because of the disk tilt in A, the angle of
attack of B at closest approach is large ($>$$80^\circ$).  Both
galaxies feel a prograde disturbance.

The tidal tails are the most dramatic structures at the later times
shown in Figure \ref{gas2}.  The complex bridge is almost as
prominent.  The same structures are also evident, though more diffuse,
in the disk stars in Figure \ref{stars}.

Figure \ref{Agas} shows three late time snapshots of just the gas
particles originating in A, while Figure \ref{Bgas} provides the
corresponding views of gas particles originating in B.  These figures
show that there is substantial mass transfer from each galaxy to the
other.  However, galaxy B loses much more mass.

Next we will examine individual collision structures in more
detail. We note at the outset that the figures illustrating these
structures have been chosen at a variety of times in the range $100 -
220 Myr$, though in most cases $t=120 Myr$. The latter figure is
essentially a default, but if the specific structure is better
illustrated by the model output at a different time, we have shown it
at that time. In some cases multiple times are shown to illustrate
structure development. The range above approximates our uncertainty in
the time since closest approach, and since the lifetime of most
structures is longer than this range, all the illustrated features
should be visible at the ``present.'' 

\subsection {Specific Collisional Structures}

\subsubsection {The NGC 7714 Ring (Feature 1)}

The high inclination orbit of galaxy B relative to the disk plane of
A, and the impact of the center of B within or near the A disk, should
give rise to a substantial m=0, ring-like collisional perturbation of
A. In optical images (Fig. \ref{arp}) a prominent stellar ring is
visible. Rings are clear in the A gas disk in Figure \ref{gas2}, but
fainter in the stellar disk. The primary reason for this is the fact
that the dissipationless star particles have much larger thermal
velocities in the model than the gas particles, so stellar waves are
heavily smoothed by particle diffusion (also see Figure \ref{sring}).

Despite this numerical complication, it remains true that this type of
collisional perturbation has strong m=0, 1 and 2 Fourier components.
The m=0 component gives rise to successive rings. The m=2 tidal
component is largely responsible for the tidal tails and two-armed
spirals in the interior, as well as the general bar-like appearance of
the disk. The m=1 and higher odd moments generate asymmetry in these
structures. All of these features are apparent in Figure \ref{sring}
which shows enlarged views of star (left panel) and gas (right panel)
particles in the galaxy A disk at a time near the present. The left
panel of the figure provides one of the best views of the asymmetric
stellar ring. Rings are more evident in the gas disk shown in the
right panel, but are rather complex structures. This is because the
rings are also spirals, which are more or less tightly wrapped at
different positions. These spirals also smoothly connect successive
rings. This is a result of having comparable m=0 and m=2
perturbations.

In fact, the tidal countertail (which corresponds to the HI loop and
Feature 2 of NGC 7714) and the companion-side tidal tail are connected
at early times. (They are also connected at late times, but the
connection is represented by so few particles, it is virtually
invisible.) Since these structures developed even before closest
approach we call them the 0 order ring. This 'ring' is very
asymmetric, with the HI loop representing the strongly positively
torqued side. Each successive ring also has both positively and
negatively torqued sides. The former (denoted by a plus sign) moves
radially outward earlier and farther than the latter (denoted by a
minus sign). At the time shown in Figure \ref{sring} ring 0 persists
because it is largely a material rather than phase wave. Much of ring 1
(a phase wave) has propagated through the disk and disappeared. Ring 2
is maturing.

Part of ring 1+ remains visible on optical images as a very faint loop
outside the bright east side ring/loop. It contributes much to the
ring-like appearance of the stellar disk in Figure \ref{sring}. It
loops inward more tightly than the corresponding gas feature, which
has a wispy appearance. We propose that the strong NGC 7714 stellar
ring is equivalent to ring 2(+ and -) in the model. The high resolution
HST (WFPC2) partial image in Figure 2 of \citet{lan01} looks very
similar to the right panel of Figure \ref{sring}, except that the 2+
ring arc is not so prominent on the west side. In the HST image there
appear to be bubbles and shells on this arc, so SF feedback may have
effected its strength and appearance.

To date, the observations do not show a strong gas ring, like that
visible in Figures \ref{gas2} and \ref{Agas}. In part, this may be the
result of limited observational resolution and sensitivity. However,
the absence of star formation in the ring also suggests that there is
not much highly compressed gas. In fact, the gas ring in Figure
\ref{sring} is weaker on the east (left) side. This is the sector of
disk-disk overlap in the collision, and also the bridge side. Large
scale shock dissipation is likely to have separated the gas and stars
in this region, with the gas either being pushed inward or pulled into
the bridge. This accounts for the fact that the east side ring
consists of old stars, with little evidence of compressed gas or young
stars.

On the other hand, the west side gas ring (2+) is strong in Figure
\ref{sring}. We have already noted that HST observations suggest SF
feedback may have heated and scattered the gas on this
side. Preliminary models with feedback also support this idea, and
further show that the 2+ ring could be completely disrupted by
reasonable amounts feedback. Observations of the stellar populations
in this region would be very desireable.

As we will discuss in Section 3.3, there is evidence that the
northeastern tail (Item 3 of Table 1) and the southern filament of
Fig. \ref{HImap} are parts of an older ring.  These may well be
remnants of the gas component to ring 1+. 

The ratio of the successive ring radii are closer to unity than in the
double-ringed Cartwheel galaxy, or most models of double rings.
Theory suggests that collisional rings are closer together in galaxies
with declining rotation curves (e.g., \citet{str90}; \citet{app96}),
and so, this provides some evidence that the halos of these galaxies
are less extensive than most.  (Paper I shows that the NGC 7714
rotation curve is flat or slightly declining, but this curve is
affected by the collisional disturbances, and doesn't extend much
beyond the optical disk.) On the other hand, the asymmetry of the
ring, and the fact that they have been torqued in the interaction,
complicates this interpretation.

\subsubsection {Outer Southwest Tail and HI Loop of NGC 7714 (Feature
2)} 

The `outer southwest tail' seen in optical images (Feature 2 of Figure
1) appears to be just the base of the great HI loop described in Paper
II (see Figure \ref{HImap}).  This stellar feature becomes broader and
fainter at a position angle of about $270^\circ$, due west of the NGC
7714 nucleus, and at low surface brightness levels traces the HI loop
to at least a position angle of $330^\circ$.

Our Hydra models (Figures \ref{gas2} and \ref{stars}, lower left
panels) also show a low surface brightness stellar feature coincident
with a gaseous loop.  As in the observations, the stellar feature is
shorter and less prominent than the gaseous feature.  This is a common
property of tidal tails (see e.g., \citet{hib96}, \citet{elm00},
\citet{mih01}); it is the result of the fact that tails are pulled out
of the disk like taffy out of a pot, with the outer boundary of the
tail derived from the outer edge of the disk, which is typically
gas-rich, while the inner base of the tail includes material from
deeper within the disk, which has a larger proportion of stars.

To help understand how such a large mass of gas ($\simeq 1.1 \times
10^9 M_\odot$) was pulled out into the huge loop, we investigated the
history of the particles in the loop in the Hydra models (Figure
\ref{loop}).  In the left panel are shown all gas particles above the
plane $y = 0.31 + 0.48x$, i.e., most of the loop material that is not
too close to the A disk. 

The right-hand panel of Figure \ref{loop} shows the location of the
same gas particles at a time shortly before the impact the two disks.
It suggests that most of the outer disk on the south side is pulled
out into the loop as the companion swings by.

However, Figures \ref{gas1} and \ref{gas2} show that while this loop has
characteristics of a classical tidal tail, it also has a strong
ring-like aspect. E.g., it is always a loop, which connects back to
the A disk, and not a tail with a detached end point. Thus, this loop
may also be regarded as part of the first ring-wave.

The fact that most of the material in the loop originates in the outer
part of the pre-collision disk also helps explain the lack of
detectable molecular material in the loop, despite the large gas mass
it contains (see \citet{smi01}).  We note that the nucleus of NGC 7714
has moderately low chemical abundances (12 + log(O/H) $\sim$ 8.5;
\citep{f80, gd95}), and the material in this loop may be even more
metal-poor than that in the nucleus.

Note that the contour levels, while arbitrary, are the same in both
panels of Figure \ref{loop}.  However, the area contained within them
is much less in the left panel, which highlights the strong
compression resulting from the collision.

\subsubsection {The Bridge (Feature 5)}

The HI maps of Paper II suggested that the bridge between the two
galaxies is a complex structure.  In particular, there is a
measureable offset between the star and HI gas centroids. To begin to
understand this let us consider the origin and development of the
bridge. A plot like Figure \ref{loop}, but for bridge particles (not
shown), tells us that most bridge particles originate in the outer
parts of their parent disks, and on the side that is closest to the
other galaxy at closest approach.

The y-z plots of Figure \ref{brorig} provide good views of the
development of several bridge components (which are labeled in the
second panel). The figure shows the results of Hydra models at two
intermediate times. The first model is the usual `best' model, while
the second is one we will refer to as the `alternate.' The alternate
model differs from the best model by having a different tilt of the B
galaxy, and higher orbital angular momentum (see Table 3). It is
representative of a group of models with slight differences from the
best model. Comparisons between the best and alternate models help us
understand which structures are common to a range of similar models,
like ring waves and the tidal tails, versus those which depend
sensitively on structural and orbital parameters. The bridge contains
components of both types.

The first bridge component (Labeled i in Figure \ref{brorig}) is the B
tidal (counter) tail, which loops back down into projection onto the
bridge in this view and especially in the x-y view. It is not
physically associated with the other (true) bridge components. The
second component (ii) is a strong tidal bridge from B to A.  This
component, the late or post-collision B bridge, begins as a dark,
nearly vertical line on the left side of the gas between the two
galaxies in the upper left panel. It also dominates the left side of
the bridge in the alternate model at this time (upper right panel).

A third component of the bridge (iii) is the tidal stream from A to
B. Bridge particles and galaxy contours are shown in Figure
\ref{bridge}. This stream generally misses its target initially in
three dimensions.  In the best model, this component does not even
stretch much towards galaxy B before it falls back to the A disk
plane. However, it remains outside the A disk and projected onto the
bridge in the x-y view through most of the simulation. This can be
seen in the upper left panel of Figure \ref{brorig}, which provides
the remaining two views of the bridge gas at intermediate times. In
the alternate model this A bridge does stretch towards galaxy B (see
Figures \ref{bridge} and \ref{brorig}). At late times it transfers gas
directly into the core of B. Figure \ref{brorig} highlights this
difference between these two otherwise very similar models. This
difference is mostly due to the different companion tilt angles. In
the best model the companion cuts through more sharply, not splashing
as much gas, just gravitationally lifting the nearby disk
slightly. Because the gas is projected onto the bridge in both cases,
observations may not distinguish between the two models in this
regard.

There is also a fourth component of the bridge (iv), which can be seen
as the sharp linear feature stretching from the B center toward the A
center in the top right panel of Figure \ref{brorig}. It is also
present in the best model, though harder to distinguish in the top
left panel of Figure \ref{brorig}.  This feature merges with the late
bridge by about $t = 100~Myr$.  Graphs at other times (not shown),
reveal that this structure is the remnant of an early or pre-collision
tidal transfer stream from B to A, which appears to transfer a
respectable amount of material (see discussion at the end of this
section). Thus, B is a remarkable case of a galaxy with a long tidal
tail, and two(!)  tidal bridges originating on opposite sides of the
galaxy. The multifaceted structure of this bridge also helps explain
how such a large mass of HI gas can be located outside the disks of
the two galaxies.

The different components of the bridge have different degrees of
gas/star offset.  The B tail has essentially no offset between stars
and gas. In Figure \ref{broff} we show two views of gas and star
particles that originate in A in the best model.  It is clear in this
figure that the stars are much more broadly distributed than the gas
particles.  It is also evident from the top panels that the center
lines of the star and gas distributions are offset. 

Moreover, the last two panels of Figure \ref{bridge} show how the
bridge gas particles from B compress into a relatively thin filament
in the Hydra models.  This compression includes a merger of the
`early' and `late' B bridges.  In the gas this process is dissipative,
but the stars are dissipationless.  This also helps explain the
offset.

From Figure \ref{brorig} (and Figure \ref{swtail} below) we deduce
that most of the bridge material is contained within tidal structures.
In Paper II we speculated that the bridge might contain a mixture of
splash and `tidal swing' material. Dissipation in the collision
between disks does seem to be an important factor in the development
of the A to B component, which is important in determining the
gas-star offset. However, this `splash' effect is more indirect than
originally envisioned. Dissipation in the later compression of the
early and late B to A components also plays a role (see also
\citet{mih01}).

These considerations suggest an unusual explanation for the young star
clusters in the bridge.  As noted, shortly before the present the late
B bridge swings east and overtakes the remnant of the early bridge.
The gas in the latter is compressed, and this compression may trigger
star formation.  The H$\alpha$ observations of Paper II suggest that
these clusters are truly young, and this is the only dynamical event
that occurs in the region within a few times $10^7~yrs.$ of the
present.  However, we do not have sufficient particle resolution in
the simulation to detect clump compression, so we cannot directly
confirm this conjecture.

Another possibility is that these star clusters are formed in the NGC
7715 tail, which experiences compression in the segment projected onto
the bridge at about the same time.  Delayed strong compression was
found to occur in many of the models of \citet{wal90}.  Kinematic
observations might allow this hypothesis to be tested.  We will
describe the kinematic predictions of the simulation below.

\subsubsection {The NGC 7715 Tail (Feature 6)}

Next we will consider the NGC 7715 (stellar and gas) countertail.  As
we have already discussed, our models indicate that this tail curves
around behind NGC 7715 and the bridge (Fig. \ref{Bgas}). Actually, in
the best model it is located a bit below the bridge in the x-y
plane. In the alternate model it is slightly above the bridge (at the
same time). The inclination of the companion disk was changed in the
best model to give a more edge-on appearance in the x-y view (see
Table 3). With a somewhat smaller change it should be possible to
obtain a quite edge-on appearance, while also superimposing the tail
on the line-of-centers of the two galaxies.

In the models, at $t \ge 100~Myrs.$, the gas in this tail extends about
25 kpc to the southwest of NGC 7714 (Figures \ref{gas2}, \ref{Bgas},
and \ref{tail}).  The long HI plume to the southwest of NGC 7714, seen
in low resolution HI maps (Figure 3b in Paper I) is very likely the
observational counterpart to this model tail.  This feature does not
have an observed optical counterpart, consistent with it being the end
of an HI-rich tidal tail. As shown in Figure \ref{tail}, almost all of
this tail material originated in an annulus in the outer part of the
NGC 7715 disk, on the side nearest the NGC 7714 disk before closest
impact.  As discussed below, the observed velocity structure of this
tail is also reasonably consistent with the model NGC 7715 tail.

\subsubsection{The Bridge Extension}

In the Arp Atlas photograph (Figure \ref{arp}), a small stubby plume
is visible on the west side of NGC 7714, exactly in line with the
bridge on the east side. This feature lies north of (above) features 2
and 3 in the figure.  It seems likely that this plume was formed by
bridge material that has fallen through the NGC 7714 disk plane.  The
model bridges do not have enough particles to confirm this, and the
available HI data do not have sufficient resolution to distinguish
this feature from the disk.

\subsubsection{The Inner Southwest NGC 7714
Tail (Feature 2)} 

The inner SW tail (Feature 3 in Figure \ref{arp}) is very unusual in
several respects.  First, it does not seem to be part of the southern
tidal tail of NGC 7714 (which is also the HI loop), but rather an
extra inner tail paralleling feature 2.  Moreover, it has a high
surface brightness, and contains knots of recent star formation.  It
also does not appear to be an extension of the bridge, nor material
accreted out of the bridge.

Figure \ref{swtail} provides a plausible solution to the mystery.  The
first panel shows how mass transfer is well underway at the time of
closest approach. Inner SW tail particles were first identified in the
third panel of Figure \ref{swtail}. That is, all gas particles from B
lying within a rectangular region between the A disk and the HI loop
were selected and marked with plus signs. This somewhat crude
procedure misses a number of the inner tail particles, and includes
some that are not truly part of that structure, and travel past it at
later times. Nonetheless, the procedure is accurate enough for
illustrative purposes. The location of the marked particles in the
first panel confirms that they are part of the earliest mass
transfer. Already by the time of the second panel ($t = 40~Myr.$) they
are part of a plume located behind the nascent HI loop.

At the time of the second panel the point of origin of the early
transfer stream is near the top of the companion disk, on the far side
relative to A, and this stream has become very thin (just to the left
of the label).  In the meantime, a second, vigorous transfer stream
has developed on the current near side of the companion.  The bridge
narrows with time, and the streams are merging by the time of the
third panel ($t = 70~Myr.$).  In this view the inner SW tail is
visible as a horizontal set of particles lying below the primary, and
above the long curved contour of the HI loop.

The final panel of Figure \ref{swtail} shows a nearly x-y view at a
time near the present, with the feature labeled.  In this plot, the
plume is offset from the tail/loop as in the observations. At this
time, this structure is in the NW quadrant, rather than the SW, but
otherwise generally matches the observations well. In the true x-y
view the plus signs lie on top of the HI loop contours. The small
rotation (about $5^{\circ}$) about the y-axis used to make the fourth
panel easily separates the two features, without greatly changing
other characteristics. Thus, the position of this feature seems to be
a very sensitive function of viewing angle. The inner SW tail is not
reproduced as well in the other Hydra models, though it is still
present.

\subsubsection {The Inner Disks and Mass Transfer}

Figure \ref{cormas} shows, as a function of time, the gas mass
contained in a spherical volume of radius 0.03 model grid units (about
3 kpc) normalized to its initial mass centered on the nuclei of the
galaxies.  The figure also shows the gas fraction in each galaxy
transferred from the other galaxy. Note that peaks in the mass
transfer at times near closest approach (t = 0) are largely due to the
proximity of the two disks, and temporary incursions into the
spherical volumes measured.

At times from t = 0 on, ring-like waves have a significant effect on
the total gas curves for the primary galaxies in all the Hydra
models. As time goes on, there is an increase in the mass of gas in
the primary core.  This is most likely due to compression resulting
from angular momentum transport in the spiral waves, and angular
momentum mixing with transferred gas.

The delayed central gas buildup after closest approach is interesting,
and agrees with the observations of delayed starbursts in interacting
systems (see e.g., \citet{ber93}).  The delay is long enough that the
tidal transfer streams are much reduced before the central buildup
gets underway.  Most of the gas transferred to A at the earliest times
falls onto the outer parts of the disk.  There is some direct transfer
from the bridge into the central regions at later times, and some
material from the outer parts is transferred inward. The transfer
history of B is more extreme.  The `bridge' from A initially misses B
almost entirely in the alternate model, and never really leaves the
parent galaxy in the best model. In the former case, gas accumulates
at the end of this bridge, and later, falls en masse onto B.

\subsubsection{Structure Summary}

Table 4 provides summary comparisons of the gas masses in different
components in the observations and the best model at a time of $170
Myr$ after closest approach. The comparisons are generally very
good. Specifically, the post-collision model disks contain gas
fractions that are quite close to those observed. (Note that about
20\% of the observed HI flux is not contained in the identified
structures, so the model figures must be renormalized to compare to
observation.) The B disk contains about twice the observed fraction,
but this is still much lower than the initial value. That is, it has
lost 2/3 of its original gas. NGC 7715 is a post-starburst galaxy, so
some of its gas may have been lost in a starburst wind, which is not
modeled. Alternately, it may simply have had less gas initially than
assumed in the model. Even though the A disk receives a substantial
amount of mass transfer, its gas fraction is also greatly reduced by
the collision.

The bridge gas fraction of the model appears to be substantially less
than observed, even including the superposed galaxy B tail. However,
this number is a very sensitive function of time and of the
post-collision deceleration. At $t = 100~Myr$ in this model, the gas
fraction in the bridge was nearly twice the value shown in Table
4. Between these two times, substantial amounts of gas falls out of
the bridge (and mostly onto the A disk). If, for example, the dark
halo of A was somewhat less concentrated than in the model, then the
galaxies would not have decelerated as quickly after closest approach,
and would have separated more, stretching the bridge, and making for a
longer fallback time.

The HI loop is a little too massive in the model, perhaps indicating
that the model A disk is too large. On the other hand, faint parts of
this diffuse structure would not be identified in the observations, so
the model result may in fact be more correct than it first appears.

\subsection{Kinematics}
\subsubsection{Velocity Maps}

In this section we compare the model and observed line-of-sight (LOS)
kinematics. Figure \ref{losvel} provides the LOS velocity map for the
multi-array VLA 21 cm observations of this system (presented in Paper
II).  In Figure \ref{modvel} we show the corresponding map for the
model. More precisely, the figure shows contours of the z-component of
the velocity across the x-y plane, and the contour values have been
scaled to be comparable to the observed values as described in the
caption. Both Figures \ref{losvel} and \ref{modvel} also show the
velocity dispersion map in gray-scale.

The observational contours in Figure \ref{losvel} are not calculated
at column densities less than a fixed minimum value, and the model
contours (and velocity dispersions) are only computed for bins
containing at least 5 particles.  The fact that these cutoffs are not
identical results in some minor differences, for example, the contours
connect across the bridge in the model plot, but not in the
observational one.  The resolution of the two plots is comparable.
The model contours are computed for data binned on a scale of 0.01
grid units, or about 1 kpc.  The observational effective beam width is
about $6''$, or about 1.1 kpc, with the assumed system distance of 37
Mpc.

The limited resolution and signal to noise in the observations (and
models) of the bridge preclude any detailed kinematic analysis there.
However, Figures \ref{losvel}, and \ref{modvel} do show general agreement
between models and observations in the bridge.  Specifically, as the
contours go from NGC 7714 across the bridge they go from fairly
horizontal (constant declination), to mixed vertical and horizontal,
and ultimately more vertical.

Similarly, in the primary galaxy (NGC 7714), the model and
observational contours are similar. However, the LOS velocity range of
Figure \ref{modvel} is slightly greater than observed. Also the model
galaxy A contours are more horizontal than the observational contours,
but they tilt more toward the vertical, and compress together, at
later times.

The agreement in the companion (NGC 7715) is also quite good, with
mostly horizontal contours in the bulk of the galaxy, which curve
upward on the bridge side in both Figures \ref{losvel} and
\ref{modvel}.  Figure \ref{modvel} shows that the velocity range is
too large in the model galaxy. This is probably the combined result of
several effects. However, because it is common to all the Hydra
models, we suspect that the dominant effect is the compact halo
distribution. 

Figures \ref{losvel} and \ref{modvel} also illustrate the spatial
distribution of LOS velocity dispersion in both the model and
observations. The velocity dispersions in the observations range up to
about 50 km s$^{-1}$, and also up to 50 km s$^{-1}$ in the model.
However, in the model the largest velocity dispersions come from cells
with only 5-10 particles, and the range is reduced by a factor of 2/3
if these cells are excluded.

In both models and observations the dispersion is larger and less
uniform in the primary (NGC 7714) than in the companion. It appears
that the distribution of velocity dispersions within the primary is
different between the models and the observations.  Observationally,
the highest dispersions are found in a double cone centered on the NGC
7714 nucleus and oriented $45^\circ$ from the vertical. In the models
the highest dispersion are found on the south side of the galaxy. It
is possible that the high dispersion values in the core of NGC 7714
are due to effects of the starburst, or an incipient wind, which are
not included in the Hydra models.  On the other hand, a significant
part of the southern dispersion in the model is due to gas accreted
from B, or in the tail.

\subsubsection{Position-Velocity Maps}

Another way to compare model and observational kinematics is by means
of position-velocity plots.  We will consider one example, the
velocity - right ascension plot.  The model results are shown in
Figure \ref{modvra}, and the observational results in Figure
\ref{obsvra}.

We have made a feature by feature comparison between models and
observations in Figures \ref{modvra} and \ref{obsvra}, and will
briefly summarize the procedure and results. Distinct features were
labeled in the observational plot (Figure \ref{obsvra}), and then the
corresponding features in the model were located on the model plot
(Figure \ref{modvra}). In most cases the correspondence was direct,
for example, the galaxy A and NGC 7714 disks are very similar.  Yet,
there are some differences. In Figure \ref{modvra} the disk of B is a
very long, and nearly vertical, line of gas at $x \simeq 0.42$. The
prominence of this feature provides yet another indication that the
velocity range of the B disk was not quite correct.

Some of the differences between these figures appears to be due to
observational sensitivity limits.  In particular, the long western HI
tail (i.e., the western extension of the ``eastern tail'' from NGC
7715, see Figure 3b of Paper I) has too a low surface brightness to
appear in Figure \ref{obsvra}, which is based on the higher resolution
HI data of Paper II. The model predicts a slightly higher velocity for
this feature ($\sim$3040 km s$^{-1}$) than the observed velocity of
the HI tail ($\sim$2850 km s$^{-1}$; Figure 4 in Paper I), however,
the sign of the velocity shift is correct, in that both the model and
the observations show that the tail is redshifted relative to NGC 7714
and NGC 7715.

At this time in the model, the Inner SW Tail largely overlaps the base
of the HI gas loop, though they were separate at earlier times.  In
the model plot (Figure \ref{modvra}), the HI loop and the eastern NGC
7715 tail appear much more extensive than in the observational plot
(Figure \ref{obsvra}). This is due to surface brightness limitations
in the observations.

The A disk is distorted in both the model and the observations, in the
sense that it does not have the form of a typical rotation curve. The
model suggests that at large x values the distortion is the result of
the tidal perturbation that is also responsible for the gas loop.  At
low x values there is mixing with accreted gas from galaxy B.

The HI maps of Paper II revealed a couple of structures on the
southern boundary of the bridge.  These are labeled the `southern
filament' and `southern arc' in Figure \ref{HImap} and in Table 1.
They do not contain a large amount of gas, and so were not discussed
above, however, they are notable as distinct features in Figure
\ref{obsvra}.

The appearance of the southern arc and filament in the HI map suggests
the possibility that they are physically connected, or at least
connected in their origin.  A careful examination of the model results
suggests that this is not the case.  The models suggest that the
southern arc is most likely part of the bridge from NGC 7714 to NGC
7715. In contrast, a close examination of the model results suggests
that the southern filament may be an extension of the 1+ ring
discussed in Section 3.2.1. If so, then the southern filament is a gas
structure corresponding to the NGC 7714 optical ring, and not, for
example, an inner branch of the HI loop.

In conclusion, the model kinematics generally agree with the
observations. In addition, dispersion measurements suggest the
possibility that the starburst in NGC 7714 has energized the gas.

\section{Star Formation History and Thermal Phases}

\subsection{Background}

One of the primary motivations for making detailed dynamical models of
individual collisional systems is to learn to what degree, and how,
collisional dynamics orchestrates large scale SF and nuclear
starbursts. By driving galaxy disks far from self-regulated steady
states, collisions provide a unique tool for studying these
processes. With good dynamical models we can hope to determine how the
SF depends on local density and pressure variations, for example.
High resolution observations are required to provide sufficient
morphological and kinematic information to constrain the
models. Multiband spectral data are needed to determine the stellar
populations present, and the SF history from observation.  Presently,
only for a few systems do we have sufficient data for a meaningful
comparison between spectral synthesis models and SF histories derived
from dynamical models.

Recently, \citet{lan01} have published an extensive spectral
synthesis model for the nucleus of NGC 7714, based on spectra
ranging from the near-IR to the UV. They explored a large range of
possible SF histories to achieve a good fit to these spectra,
including models with a continuous component of the SF, and up to
three distinct bursts.

Unfortunately, despite an impressive number of spectral constraints,
they found that the model fits were not unique.  In particular, they
found that two very different SF histories could account for the
observations equally well. The first history consisted of three
starbursts, with the earliest occuring about 500 Myr. ago, and the
other two occuring at times of about 20 and 5 Myr. before the
present. The amplitude of these bursts declined with time. The second
model consisted of a 5 Myr. burst, plus a continous SF component,
which began about 300 Myr. ago, and declined in amplitude since that
time.  In both these models, a very old stellar component is also
assumed to be present.  The conundrum emphasized by Lancon et al., is
that both models require either a burst or the onset of a significant
continuous component at a time before the current collision. We shall
take this question up again shortly, but first, we consider what SF
histories are suggested by the dynamical models presented above.

\subsection{Model Results}

We have run versions of the model with feedback heating from star
formation, and the results of these models are in qualitative
agreement with the observational results. E.g., the net star formation
of the system is dominated by that in the central regions of galaxy A;
the star formation in outer disk waves and tidal structures is
generally small or absent. Secondly, the star formation within the
center of galaxy A frequently has a burst-like nature, with 2-3 bursts
within the time since closest approach. This is encouraging as far as
it goes, but it is beyond the scope of this paper to quantify these
results and make more detailed comparisons to observation.

There are two main reasons for this limitation on the results. The
first is that feedback models introduce several new parameters, e.g.,
density and temperature thresholds for the onset of feedback, the
amount of energy input per feedback event, time scale for the energy
input, and a refractory time scale for the affected gas particles
after feedback. The consequences of the feedback depend on timescale
ratios involving these and other intrinsic parameters (e.g., cooling
times and local dynamic times). To evaluate the validity of specific
quantitative results of simulations with feedback, we need a number of
runs to explore the region of the parameter space appropriate to the
system being modeled. (Moreover, this is an indirect problem because
it requires knowledge of the pre-collision gas distribution in the
galaxies.)

A more fundamental difficulty is that feedback models which give
post-collision bursts, and in which the gas distribution in the galaxy
cores is not greatly changed by the collision (as in this system), are
generally also bursty before the collision. The timing and size of the
post-collision bursts then depend to some degree on the nature of the
last precollision burst(s). As we plan to show in a separate paper, we
believe that such models correctly capture the intrinsically recurrent
nature of central starbursts in late time galaxy disks. In isolated
galaxies, such recurrent starbursts will ultimately redistribute,
heat, or exhaust the cold gas, until the core is below a critical
(density) threshold. (Although this latter threshold may simply be one
such that below it, the on-going star formation rate is too low to
generate significant feedback effects.) Mass transfer or collisional
compression may push the density above the threshold again, and set
off a new cycle of recurrent starbursts. The fine-tuning of the
pre-collision conditions to simulate this is also beyond the scope of
this paper.

If the cores of the pre-collision galaxies in this system were below a
feedback threshold, then their star formation behavior will be
sensitive to changes in the central gas density and pressure. This
conjecture takes us back to Figure \ref{cormas}, which shows central
gas density in the model.  The figure shows a steady gas density
buildup in the core of galaxy A, along with small density bursts.  The
bursts are associated with the formation of an asymmetric ring
waves. These waves may induce starbursts if the gas is near threshold
and not in a refractory period. The monotonic growth of gas density
with time would make us expect strong bursts eventually, even without
wave triggers. The onset of such bursts is probably strongly influenced
by the recovery from feedback effects, assuming the mass of gas throw
out in burst-driven winds is not too great. If not, waves and recovery
processes (like cooling) probably interact with each other. These
processes will be difficult to model in detail.

Figure \ref{cormas} shows that the companion (B) center experiences a
`density' burst at the time of closest approach, which is considerably
stronger than that experienced by A. This is largely due to the
overlap with the A halo at that time, and the resulting compression.
However, the disk of B also rings, even more vigorously than that of
A.

Almost all of our models show that a significant amount of gas is
removed from the companion disk in the collision, and accreted onto
A. As discussed above, some of this material probably ends up in the
inner SW tidal tail (Feature 3), perhaps triggering the observed SF.
The rest ends up in an accretion disk, which is appears only slightly
tilted with respect to the original primary gas disk, and roughly
$2/3$ as large. 

This may account for some of the diffuse {$H\alpha$} emission observed
in the NGC 7714 disk (Paper II and references therein), which extends over
a similarly sized part of the disk.  The mid-infrared emission
detected in ISO observations \citep{oha00} also has a similar extent.
O'Halloran et al. suggest that the source of this emission is
hydrogenated, but not highly ionized, PAHs, and in the case of a 
$9.6~{\mu}m$ line, molecular hydrogen that could be shock excited.

The ROSAT X-ray observations of \citet{pap98} show two sources in NGC
7714: the starburst nucleus, and a second source that they tentatively
identify with either a wind outflow or an accretion stream out of the
bridge. If the latter interpretation is correct, the second source may
be a hot spot where the stream has hit the more relaxed material in
the disk. Our models suggest that this stream would be quite weak and
of limited extent by the present time. This is qualitatively
consistent with a `hot spot,' rather than a large hot region. On the
other hand, our feedback models also produce a vertical expansion
(like a weak wind) following starbursts, so the models favor neither
cause.

One important caveat is that the HST imagery and GHRS observations
modeled by Lan\c{c}on et al. resolve the inner 330 pc. of the nucleus,
while our simulations do not resolve scales much smaller than about
1.0 kpc. Both the smallness and the incommenserability of these scales
is a problem. Given the nonlinearity of feedback effects, and their
dependence on many variables, they can easily induce small scale
bursts, which will seem like essentially random occurances. Thus, data
over a larger region are needed before we can compare models and
observations with confidence, and so, despite the wealth of data
available, we cannot yet do so in this system.


\section{Summary and Further Questions}

Because of its proximity and its rich but not yet relaxed tidal
structure, we are able to reconstruct the collisional history of NGC
7714/5 in detail. The models presented here have collisional
parameters that are qualitatively similar to those of the earlier
models of Papers I and II. The models and the observed morphologies
suggest a recent collision ($\sim100-200~Myr$ ago). Both our new
models and previous models require an orbit of at least moderate
inclination relative to the primary disk, and a center of impact in
the outer disk of the primary, but beyond that there are significant
differences.  The most important of these are the orientation of the
companion disk, and the fact that for the primary the collision has a
significant prograde component to the perturbation (as well as
orthogonal). In the earlier works, this perturbation component was
retrograde. While these earlier models did succeed in producing both
ring and tail structures in the primary, these did not match the
observed structures nearly as well as the present models.
  
The models described above do reproduce most of the observed
morphological and kinematic structure of the system, including the
following specific features.  (Also see Table 4 for a summary of gas
masses contained in the collisional components.)

1) {\it{Rings}} (Feature 1 of Paper I) in the primary disk are
naturally accounted for by the near central impact and the moderately
high inclination of the companion's orbit relative to that disk. The
stellar rings are weak in the Hydra models, but this is most probably
because the initial stellar velocity dispersion was high. (The NGC
7714 stellar disk is also likely to be more massive than in the
models, see Section 2.3.) The prominent ring of the optical images
consists primarily of old stars, without the numerous young star
clusters that characterize other gas-rich collisional rings like the
Cartwheel \citep{hig95}.  We can suggest a couple of effects that
might be responsible. The first is that much of the gas in the outer
disk is flung out in tails and the bridge, and so, if the ring is near
the outer edge of the remnant disk, there may be little gas left
there. A related factor is that the collisional overlap with the
companion disk was on this side of the primary disk. Secondly, the
radial perturbation is not as great as in a classical ring galaxy, so
gas compression in the wave is less. The models suggest that this
strong asymmetric ring may be the second ring wave. Traces of the
first may be seen as a faint (east side) feature on deep optical
images, and perhaps in the gas as the ``southern filament.'' 

2) {\it{The HI loop and the optical SW tail}} (Feature 2) are parts of
the same tidal structure, which is a tidal countertail, produced by
the prograde component of the disturbance. Although it originated as a
tidal tail, the HI loop is ringlike, and is described above as a
precursor or zeroth order ring.

3) {\it{The connecting bridge}} (Feature 5) is predicted to consist of
multiple components: early and late forming bridges from NGC 7715, a
bridge pulled from NGC 7714, and the superimposed tidal countertail
from NGC 7715. The superposition of these components accounts for the
large gas mass of the bridge. 

4) {\it{The tidal tail}} originating on the east side of NGC 7715
(Feature 6) may curve behind the disk of its parent galaxy and the
bridge, and appear on the far west side of NGC 7714 as an extended HI
tail. The models suggest that this is a long and massive feature,
though largely hidden. More generally, they suggest that much of that
galaxy's original disk has been ripped out. This helps explain the low
gas mass in NGC 7715.

5) {\it{The stubby NW plume}} of NGC 7714 is most likely an extension
of the NGC 7715 (late) bridge. This conclusion is based mainly on
optical imagery, as this feature is not clearly resolved in the HI
maps. This feature is minor, and the models do not have sufficient
resolution to confirm it as a bridge extension.

6) {\it{The inner SW tail}} (Feature 2) of NGC 7714 is one of the most
mysterious structures in this system.  The Hydra models suggest that
it is the result of material transferred at early times from the
companion. It was subsequently torqued out into a tail, like other
material in the NGC 7714 disk, but at an slightly different location.

7) The models suggest that {\it mass transfer} onto the NGC 7714 disk
has been prolonged and significant, but there has been little
accretion onto NGC 7715 as yet.  (This result should be treated with
some caution, since it depends on the specific orientations of the two
disks.)

8) There is general agreement between model and observed (HI)
kinematics, including the distributions of both line-of-sight
velocities and velocity dispersions, however, there are several
differences in detail. E.g., the western HI tail is somewhat more
redshifted in the model than in reality.

9) We achieved better fits to the collisional morphologies with models
with a halo potential that yields a flatter rotation curve than the
softened point mass potentials used in the earlier models.
Nonetheless, these latter halos diminish quickly at radii greater than
the initial disk radius, in accord with the criterion of \citet{dub99}
for halo structure in tailed systems.

These comparisons show that the simulations above have provided one of
the most successful and detailed collisional models of a disturbed
galaxy system to date.  This model reconstruction not only provides an
example of the power of collisional theory, but it also provides a
basis for studying SF on a region by region basis within these
systems.  It makes it possible to test our understanding of the
mechanisms of interaction-induced star formation, and the role of
various gas dynamical processes, with much more precision than would
be possible without such information.  We obtained the following
specific results.

10) {\it{The SF history in the center of NGC 7714}} is driven, in
part, by multiple ring compressions. Multiple bursts have also been
suggested by the population synthesis models of \citet{lan01}.  These
synthesis models further suggest that an episode of SF began a
considerable time before closest approach (i.e., $100-400~Myr$
earlier).  Our numerical simulations do not resolve the mystery of how
this SF might have been caused.  They suggest the possibility that
compression began somewhat before closest approach. Since the time
since closest approach could be quite long (up to $200~Myr$) in an
encounter like that of the best model, triggering by this early
compression might suffice to explain the early burst. Another
possibility is triggering by an earlier encounter, even if this
encounter was distant and the perturbation small. The relative orbit
of the galaxies, at a time much before closest approach, is not well
constrained.

11) {\it{NGC 7715 is characterized as a post-starburst galaxy.}} The
models yield a compression at the time of closest approach that might
be expected to trigger a burst. From that time to the present the loss
of disk material, and lack of mass input, result in the suppression of
SF in this disk.

12) The models provide intriguing hints about regional SF.  For
example, the model result that the bridge consists of multiple,
generally non-overlapping and low density parts, helps us explain the
apparent contradiction between the large gas mass but low SFR in the
bridge.  At the same time, there is a beautiful filament of SF knots
on the north side of the bridge.  Our models suggest that this SF was
triggered by an interaction between bridge components.

Another interesting SF region is the inner SW tail. Our models suggest
that this is a mixing region between disk and accreted material. Thus,
SF may be the result of turbulence and enhanced compression.

These results on SF are tentative, because the resolution of
compressed regions is limited, and the representation of thermal
processes is approximate.  Yet the qualitative agreement between the
dynamical models and spectral evolutionary models for the recent SF
history of this system is encouraging.  At the very least, these
results yield no contradiction to the proposition that
collision-induced compression drives the SF.

To date, there have been few cases in the literature where the
spatial/temporal pattern of compressional disturbances is
quantitatively compared to that of its young to intermediate age
stellar populations.  The extreme case of large amounts of gas dumped
into the cores of major merger remnants, followed by super-starbursts,
provides one example.  Classical collisional ring galaxies like the
Cartwheel, with aging burst populations behind the propagating wave
provide another example (see \citet{app96}). Our success with this
system offers hope that we can also succeed in other, less simple,
systems. We need many such comparisons in order to advance our
understanding of how interactions induce SF.

Despite the fact that this system has been observed in almost every
waveband from radio to X-ray with a wide variety of ground and
space-based telescopes, it remains true that additional observations
would be helpful.  Increased resolution of the distributions of gas
and stellar populations would be very helpful for answering the
remaining questions, and guiding more refined models. For example, the
HI observations of Paper II barely resolve important structures like
the ring and the bridge.  As a second example, extensive HST imaging
and spectral observations have been made of the starburst core of NGC
7714 (see section 1), with snapshots of various parts of the
system. However, no complete HST imaging survey has been undertaken.
High quality spectral observations of SF regions outside the core
would also be very useful.  When higher resolution observations become
available, the simulations presented here could serve as a starting
point for a new generation of models with much greater particle and
spatial resolution.

\acknowledgments

This research is based in part on archival 21~cm HI observations made
in 1989, 1992, and 1994 with the National Radio Astronomy Observatory
(NRAO) Very Large Array (VLA) telescope, a facility of the National
Science Foundation, operated under cooperative agreement by Associated
Universities, Inc.  This research has made use of the NASA/IPAC
Extragalactic Database (NED) which is operated by the Jet Propulsion
Laboratory, California Institute of Technology, under contract with
the National Aeronautics and Space Administration.  We have also made
use of the Digitized Sky Survey, a compressed digitized form of the
Palomar Observatory Sky Atlas, which was produced at the Space
Telescope Science Institute under US Government grant NAGW-2166.  The
National Geographic Society-Palomar Observatory Sky Atlas (POSS-I) was
made by the California Institute of Technology with grants from the
National Geographic Society.  We thank John Wallin for his restricted
3-body code, which we used for preliminary modeling runs.

\clearpage
 
\begin{deluxetable}{lcc}
\scriptsize
\tablecaption{Morphological Characteristics of the NGC 7714/15 System.
\label{tbl-1}}
\tablewidth{0pt}
\tablehead{
\colhead{Feature} &
\colhead{Notes} &
\colhead{References}
} 
\startdata
1. (1) Stellar Ring\tablenotemark{a} &NGC 7714 &1\\
&(No gas ring) &3\\
2. Stellar Bar &NGC 7714 &2\\
3. (4) NE Stellar Arm &NGC 7714 &1\\
&(No gas counterpart)\\
4. (3) Inner SW Stellar Tail &NGC 7714 &1\\
5. (2) Outer SW Stellar Tail &NGC 7714 &1\\
&(Associated with \\
&gas loop)\\
6. Large Gas Loop &NGC 7714 &3\\
& (NW - NE)\\
7. Edge-on Shape & NGC 7715 &1\\
8. (5) Stellar Bridge &&1\\
9. (5) Gas Bridge 
&High column density HI,&4\\ 
&offset to the north\\
&of the stellar bridge\\
10. (6) Stellar Countertail &NGC 7715 &1\\
11. Gas Countertail &NGC 7715 &3\\
12. Far West HI Clouds &In low resolution &4\\
&HI observations\\
13. Southern Arc &NGC 7714 &3\\
14. Southern Filament &NGC 7714 &3\\
\enddata


\scriptsize 
\tablenotetext{a}{Number in parenthesis is feature number 
in ref. 4}
\tablenotetext{1}{Arp 1966}
\tablenotetext{2}{Bushouse \& Werner 1990}
\tablenotetext{3}{Smith, Struck \& Pogge 1997}
\tablenotetext{4}{Smith \& Wallin 1992}
 
\end{deluxetable}

\clearpage

\begin{deluxetable}{lcc}
\footnotesize
\tablecaption{Other Characteristics of the NGC 7714/15 System.
\label{tbl-2}}
\tablewidth{0pt}
\tablehead{
\colhead{Feature} &
\colhead{Description} &
\colhead{References}
}
\startdata
\\
\multispan{3}\hfill Kinematic Features\hfill\\
\\
15. Mean Radial Velocities\tablenotemark{a} 
&Similar in &3\\
&NGC 7714 \& NGC 7715\\
16. Bridge Radial Velocity &Blueshifted relative &3\\
&to galaxies\\
17. ``Spider Diagram'' &E.g., line of nodes &3\\
&\& rotation sense in NGC 7714\\
18. Rotation Curve &NGC 7715 &3\\
\\
\multispan{3}\hfill Star Formation Features\hfill\\
\\
19. Central Starburst &NGC 7714 &2 \\
20. Arc of Knots in Bar &NGC 7714 &2\\
21. Inner SW Arm &NGC 7714 &2\\
22. Post-starburst Center &NGC 7715 &5\\
23. Bridge Knots &Several dozen &3\\
\enddata

\scriptsize
\tablenotetext{a}{Feature numbering continued from Table 1.}
\tablenotetext{1}{Arp 1966}
\tablenotetext{2}{Bushouse \& Werner 1990}
\tablenotetext{3}{Smith, Struck \& Pogge 1997}
\tablenotetext{4}{Smith \& Wallin 1992}
\tablenotetext{4}{Bernl\"ohr 1993}

\end{deluxetable}

\clearpage

\begin{deluxetable}{lcc}
\footnotesize
\tablecaption{Model Parameters.\tablenotemark{a}
\label{tbl-3}}
\tablewidth{0pt}
\tablehead{
\colhead{} &
\colhead{Primary Galaxy} &
\colhead{Companion} 
}
\startdata
\underbar{Initial Galaxy Parameters}&&\\
\\
Masses\tablenotemark{b}   ($M_\odot$):&&\\
Halo& $6.0 \times 10^{10}$ & $2.0 \times 10^{10}$\\  
Gas Disk& $2.9 \times 10^{9}$ & $0.97 \times 10^{9}$\\
Stellar Disk& $2.9 \times 10^{9}$ & $0.97 \times 10^{9}$\\
\\
Radii (kpc):&&\\
Halo & 12. (core = 2.4) & 6.0\\ 
Gas Disk&10.&7.5\\
Stellar Disk&7.5&6.0\\
\\
Disk Peak Rotation (km/s)& 200. & 200.\\
Disk Orientation Angles\tablenotemark{c}& $30^{\circ}$ about
y-axis& $-10^{\circ}$ [$30^{\circ}$] about\\
&&the z-axis, then \\ 
&&$-20^{\circ}$ [$40^{\circ}$] about\\
&&the y-axis\\   
\\
\\
\underbar{Orbital Parameters}\\
\\
Initial Center Positions\tablenotemark{d}& (12.6,
-5.6, 19.2) & (-12.6, 5.6, -19.2)\\ 
 ((x, y, z) in kpc) & [11.5, -7.0, 17.5]&
[-11.5, 7.0, -17.5]\\
\\
Initial Center Velocities\tablenotemark{e} & (-78.0, 22.5, --71.1)&
(78.0, -22.5, 71.1)\\
 (vx, vy, vz in km/s)& [-74.2, 19.5, -54.7]&
[74.2, -19.5, 54.7]\\
\\
Gas Disk Center Positions& (-2.2, 0.4, 0.0)& (1.8, -0.1, 0.0)\\
 (at Closest Approach) &[-3.3, -0.2, 0.0] &[3.2, 0.3, 0.0]\\
\\
Gas Disk Center Velocities \tablenotemark{f}& (18.6, 39.1 -229.)& (-112.,
-107., 234.)\\
 (at Closest Approach) &[0.0, 88., -180.] &[0.0, -78., 166.]\\
\enddata

\scriptsize

\tablenotetext{a}{For the best model, with values for the alternate
model in square brackets. Physical units used here.  Conversion from
code units described in text}
\tablenotetext{b}{In the self-consistent Hydra models the number of
particles in the primary galaxy halo, gas disk and stellar disk are
10,000, 9550, and 9550, respectively.  In the best and alternate
models the corresponding numbers in the companion are 3000, 3450,
3450. In all of the model galaxies the mass of a halo particle is $6.0
\times 10^6 M_\odot$, and the mass of a star or gas particle is $3.0
\times 10^5 M_\odot$.}
\tablenotetext{c}{The primary galaxy is initialized in the x-y plane,
which is taken as the fundamental plane, or the plane of the sky. The
disk is then rotated as described.  The companion galaxy is
initialized in the x-z plane.}
\tablenotetext{d}{The initial separation is 44.2 kpc.}
\tablenotetext{e}{Relative velocity is 188 km/s.}
\tablenotetext{f}{Net relative velocity is 380 km/s, with significant
uncertainty due to distorted centers.}

\end{deluxetable}
\clearpage

\begin{deluxetable}{lcc}
\footnotesize
\tablecaption{Gas Fractions in Different Structural Components.
\label{tbl-4}}
\tablewidth{0pt}
\tablehead{
\colhead{Feature} &
\colhead{$HI + {H_2}$ Mass\tablenotemark{a}} &
\colhead{Percent of Total Gas}
}
\startdata
\\
\multispan{3}\hfill From Observation \hfill\\
\\
NGC 7714 Disk&$1.7 + 2.2$&42{\%}\\
NGC 7715 Disk&$0.31 + (\le 0.13)$&3.4{\%}\\
\\
NGC 7714 HI Loop&$1.1 + (\le 0.26)$&12{\%}\\
The Bridge (Feature 5)&$1.5 + (\le 0.094)$&16{\%}\\
NGC 7715 East Tail&$0.38 + (\le 0.061)$&4.1{\%}\\
(Feature 6) &&\\
\\
{\bf{Totals}}&\bf{$7.0 + 2.2$}&\bf{77.5{\%}}\\
\\
\multispan{3}\hfill From the Hydra Model (With 1:3 Mass Ratio
 at t = 170 Myr)
\hfill\\ 
\cutinhead{Gas Mass (No. of particles)}
\\Pre-Collision Disks: A&9550&73{\%}\\
~~~~~~~~~~~~~~~~~~~~~~~~~~~B&3450&27(\%)\\
\\
Present Values (t = 170 Myr):&&\\
Galaxy A Disk&6906~(= 6064 + 842)\tablenotemark{b}&53{\%}\\
Galaxy B Disk&1000~(= 5 + 995)&7.7{\%}\\
\\
Galaxy A Loop&$\simeq 2873~(= 2747 + 126)$&22{\%}\\
Bridge (From A)&452&3.5{\%}\\
Bridge (From B)&203&1.6{\%}\\
Galaxy B Tail&1033&8.0{\%}\\
B Tail on Bridge\tablenotemark{c}&129&1.0{\%}\\
B Tail East\tablenotemark{d}&339&2.8{\%}\\
\\
\bf{Total}&&\bf{99.6{\%}}\\
\enddata

\scriptsize

\tablenotetext{a}{HI data from Smith, Struck \& Pogge 1997 and $H_2$
data from Smith \& Struck 2001.  All masses in units of $10^9
M_\odot$.  The molecular gas masses have been calculated assuming
the standard Galactic N$_{H_2}$/I$_{CO}$ conversion factor
(Bloemen et al. 1986).}
\tablenotetext{b}{The numbers in parenthesis indicate contributions
from Galaxy A and B, respectively.}
\tablenotetext{c}{That is, the part of the Galaxy B counter-tail
superimposed on the bridge in the x-y view.}
\tablenotetext{d}{That part of the B counter-tail that is east of the
B disk in the (x-y) projection.}

\end{deluxetable}

\clearpage

\clearpage

{\bf Captions}

\figcaption{The Arp Atlas (1966) image of NGC 7714/5.  North is up and
east to the left.  NGC 7714 is the larger galaxy to the west.  The
field of view is 5\farcm0 $\times$ 3\farcm9. \label{arp}}

\figcaption{ A broadband red (F606W filter) Wide Field Planetary
Camera 2 (WFPC2) image of NGC 7714 and the NGC 7714/5 bridge, from the
Hubble Space Telescope archives.  The field of view is 2\farcm5
$\times$ 1\farcm5.  The image has been mosaicked and rotated such that
north is up and east is to the left.  NGC 7715 lies off the image to
the east.  Note the prominent H~II regions between the two galaxies,
offset to the north of an optical continuum bridge. \label{hst}}

\figcaption{The VLA naturally-weighted HI map from Paper II
(greyscale), superposed on the Digitized Sky Survey (DSS) POSS-I red
image (contours).  The resolution of the HI map is 11\farcs02 $\times$
8\farcs48 with a position angle of $-$36.6$^{\circ}$, while the DSS
image has been smoothed to 12$''$ resolution.  Various features have
been marked.
\label{HImap}}

\figcaption{Gas particles in the galaxy disks at three early times in
the best model collision, with three orthogonal views at each time.
The first row shows the initial disks ($t = -180~Myr.$). The second
row shows the disks at the onset of the collision ($t = 0.0~Myr.$),
and the last row shows the disks at a time just after the closest
approach of galaxy centers ($t = 40.0~Myr.$).  The labels A,B in this
and the following figures denote the model galaxies representing NGC
7714 and NGC 7715, respectively. For galaxy A every third gas particle
is plotted, for B every particle is plotted.  Coordinates are given in
dimensionless code units, which can be scaled as described in the
text. (Note: we assume that the observer is located at the appropriate
distance along the negative z axis.)
\label{gas1}}

\figcaption{Gas particles from the galaxies, as in the previous
figure, but at times after closest approach ($t = 120,~170,~and~220~
Myr.$). This sequence illustrates the development of tidal tails and
the collisional bridge. \label{gas2}}

\figcaption{Three views of collisional distortions of the stellar
particles at the same times as in the previous figure for the gas
particles.  The late time x-z view shows that part of the stellar
`bridge' is essentially a near side tail (at the top) in this model.
The absence of an object B counter tail in the late time x-y view
(lower left) also shows that it is superimposed on the bridge (see
Figure \ref{gas2}).\label{stars}}

\figcaption{Same as Figure \ref{gas2}, but including only gas
particles that originated in object A (model NGC 7714).\label{Agas}}

\figcaption{Same as Figure \ref{gas2}, but including only gas
particles that originated in object B (model NGC 7715).\label{Bgas}}

\figcaption{{\it{left:}} The distribution of stars, and a (weak)
stellar ring in the primary disk at a time near the present in the
best model.  {\it{right:}} same for gas particles. The dynamically
cooler gas disk shows the ring waves more clearly. Several of these
are labelled as described in the text.  \label{sring}}

\figcaption{Gas particles in the best model representing the NGC 7714
`HI loop' or tail (Feature 2 of Paper I) are shown in the left panel
at time $t = 120~Myr.$, as well as a few representative contours to
mark the location of the galaxy disks.  The right panel shows the same
gas particles at a time ($t = 0.0~Myr.$) just before closest approach,
and the sample disk contours. At times between those of the two panels
the companion moves clockwise around from the right to the left and a
bit below the primary. \label{loop}}

\figcaption{Y-Z views at two intermediate times provide a good
perspective on the origin of the bridge particles in the best and
alternate Hydra models.  Gas particles from object B are shown as
dots, while the gas distribution of A is shown by a few representative
column density contours.  The bridge in the alternate model is made up
of four contributions (labeled i-iv). Features ii and iv are the dense
streams that make up the outer edges of the broad bridge from B to
A. In the best model the projected B-to-A bridge is much narrower,
though of comparable mass.  The third (iii) is the tidal bridge from
A, visible in the contours at the earlier time. The fourth is top and
end portions of the B tail that happen to be projected onto the bridge
in the x-y view.  See Fig. \ref{swtail} and text for more
details.\label{brorig}}

\figcaption{Two views of gas particles in the bridges of the best and
alternate models at t = 120 and 100~Myr., respectively. Membership in
the bridge is determined primarily by x-z position (because the bridge
stretch is large in that view). The x-y views (first column) of the
two models are quite similar. The x-z views show that in the best
model the gas particles from A promptly fall back to the left hand
side of the A disk, while in the alternate model these particles are
spread across the bridge. \label{bridge}}

\figcaption{Two orthogonal views each of the stars (first column) and
gas (second column) in the best model at a time $t = 140~Myr.$ Gas
particles from object A are shown as dots, while the gas distribution
of B is shown by a few representative contours.  The plots in the
first row show that the primary's gas bridge is concentrated on the
north (top) side of the tidal bridge, while the old stars are more
smoothly distributed. This shows the gas/star offset in this component
of the bridge. The gas tail from B (not shown) is located in true thin
filaments, stretching between the two galaxies. The panels in the
second row show that this is not the case for the gas bridge from A,
which remains near the A disk.\label{broff}}

\figcaption{Gas particles in the model NGC 7715 tail are shown in the
left panel ($t = 120~Myr.$), with representative contours to mark the
location of the galaxy disks.  All gas particles with coordinate
values $z \ge 0.63$ or $x \ge 0.37$ were included.  Most tail
particles satisfy these constraints, while particles in the disks and
other components do not (see Figure \ref{Agas}).  As in Figure
\ref{brorig} the right panel shows the same gas particles at a time
($t = 0.0~Myr.$) just before closest approach in the x-z view.  It
suggests that a large fraction of the outer NGC 7715 disk is pulled
out in the tail (26\% of all gas particles). Note that the B galaxy
contours at the earlier time primarily show the dense mass transfer
stream; the B disk is shown in part by the dots. \label{tail}}

\figcaption{Four snapshots illustrate the origin of the inner SW tail
in NGC 7714 as a result of early mass transfer from NGC 7715.  Gas
particles from the companion are shown as dots, while the gas
distribution of the primary is shown by a few density contours. Most
of the inner SW tail particles are marked with plus signs.  The plus
sign particles were identified on the basis of their position in the
third panel. The final panel shows a slightly rotated x-y view at a
time slightly before the present (see text). The inner SW tail is
labeled. The HI loop is marked by clumpy contours extending out and
upward from the right side of the primary.  Most of the particles at
the bottom of this panel originate in the B tidal tail.
\label{swtail}}

\figcaption{Gas mass relative to the initial value contained within a
spherical volume of radius $r = 0.03$ grid units (about 3 kpc.) about
the centers of the Hydra model galaxies as a function of time. The
solid curves are for the galaxies in the best model (galaxies A \& B
labelled).  Dotted curves show the fractional gas mass around each
center that was transferred from the other galaxy in the model.  The
curve with plus signs (+) shows mass transferred to galaxy B from A,
while the curve with open circles shows transfer to A from B. The
vertical dotted line shows the approximate time of closest
approach.\label{cormas}}

\figcaption{Observational LOS velocities (contours) and dispersions
(greyscale), from the data in Smith et al. 1997, for comparison to the
model results.  The contour interval is 10 km s$^{-1}$.  Some contours
are labeled.  The range in dispersion plotted extends from 0 km
s$^{-1}$ to 50 km s$^{-1}$, with darker grey corresponding to higher
dispersions.\label{losvel}}

\figcaption{X-Y views of the best model kinematics. Contours show
line-of-sight velocities, where model velocities have been scaled to
the observed range by multiplying by the scale factor of 100 km/s, and
adding 2800 km/s (the approximate systemic velocity). The local
velocity dispersion is shown in gray-scale, with regions of high
dispersion shown darker, and with a linear scale ranging from about 5
to 50 km/s. The outer perimeter contours of this plot are not
significant; this plot was made using Matlab, which requires closed
contours.
\label{modvel}}

\figcaption{Velocity of model gas particles in the z direction versus
x coordinate value for comparison to observational velocity-right
ascension plot.  Note that we have rotated the model results by
$-10^\circ$ in the x-y plane, so that the x-axis of the model more
nearly corresponds to right ascension.  The model velocities have been
scaled as in Figure \ref{modvel} by multiplying by a factor of 100,
and adding a constant systemic velocity of 2800 km/s.  The x-axis
values are in dimensionless code units. Text labels identify particle
sets corresponding to those of the observational plot (Figure
\ref{obsvra}). See text for details.\label{modvra}}

\figcaption{Observational velocity-right ascension plot.  Individual
features are marked.  The `southern filament' is a stream of gas
extending to the south of the bridge, while the `southern arc' is gas
to the southeast of the NGC 7714 disk (see Paper II and Figure
\ref{modvra}).  Note that in both of the NGC 7714 and NGC 7715 disks,
the velocities tend to increase from east to west, as does the gas in
the northwest loop and in the bridge.  The velocity of the gas in the
inner southwest tail, however, decreases from east to west.  As noted
in the text, the long southwestern HI tail seen in low resolution HI
maps (Smith $\&$ Wallin 1992) has too low surface brightness to be
visible in this plot.  It extends to RA 23$^{\rm h}$ 33${\rm m}$
15$^{\rm s}$ with a velocity of $\sim$2850 km s$^{-1}$ (Smith $\&$
Wallin 1992).\label{obsvra}}

\end{document}